\documentstyle[11pt,aaspp4]{article}
\slugcomment{Accepted by PASP: 19 May 2000}

\begin{document}
\def\wisk#1{\ifmmode{#1}\else{$#1$}\fi}
\def\um     {\wisk{{\rm \mu m}}}
\def\etal   {et~al.~}
\def\deg    {\wisk{^\circ}}
\def\icm    {\wisk{\rm cm^{-1}}}
\hfuzz=10pt \overfullrule=0pt
\pretolerance=10000
\raggedright
\pretolerance=1000	

\title{Cosmic Ray Rejection and Readout Efficiency for Large-Area Arrays}

\author{D.\ J.\ Fixsen\altaffilmark{1}, J.\ D.\ Offenberg\altaffilmark{1}, 
R.\ J.\ Hanisch\altaffilmark{2}, J.\ C.\ Mather\altaffilmark{3}, \\ 
M.\ A.\ Nieto-Santisteban\altaffilmark{2}, R.\ Sengupta\altaffilmark{1}, and
H.\ S.\ Stockman\altaffilmark{2}} 
\altaffiltext{1}{Raytheon ITSS}
\altaffiltext{2}{Space Telescope Science Institute}
\altaffiltext{3}{NASA Goddard Space Flight Center}


\begin{abstract}
We present an algorithm to optimally process uniformly sampled array image 
data obtained with a nondestructive readout. The algorithm discards full 
wells, removes cosmic ray (particle) hits and other glitches, and makes a 
nearly optimum estimate
of the signal on each pixel. The algorithm also compresses the data.
The computer requirements are modest, and the results are robust.
The results are shown and compared to results of Fowler sampled and processed 
data. Non-ideal detector performance may require some additional code,
but this is not expected to cost 
much processing time. Known types of detector faults are addressed.

\end{abstract}

\section{Introduction}
The optimal rejection of cosmic ray glitches from astronomical images is of 
critical importance for large-area pixelized detectors in space. The 
detectors (CCDs, etc) are generally stable and repeatable, so they can be 
carefully calibrated. They are often sensitive to cosmic rays and other 
radiation. The signals from cosmic rays can be the largest contamination. 
But the contamination is far from Gaussian; it tends to be dominated by ``glitches" 
which have a large effect on one or a few pixels for a short duration. Finding, 
limiting, and rejecting affected data is a problem common to many observation 
and data reduction strategies.

Cosmic rays affect ground-based detectors as well, but to a much smaller extent.
This is partly because the atmosphere and magnetic field act as a shield, 
eliminating most of the cosmic rays, and partly because atmospheric emission 
and scattering are variable and limit the extent to which it is possible to
uncover and understand other systematic errors.

This study was focused on the Next Generation Space Telescope
(Stockman \etal\ 1998); however, many
of the results are applicable over a wide range of observatories. Although
we specifically consider IR detectors, the results can be applied to visible 
light detectors or any detectors where both readout noise and Poisson statistics
are present.

\section{Noise Estimation} 
The ultimate limit on the signal-to-noise ratio is the Poisson variation in the 
photon arrival rate. For large numbers of photons, $N$, and low photon 
occupation numbers, the variance is approximately 
equal to the number of photons, so the signal-to-noise ratio is $\sqrt N$ if other 
sources of noise are negligible. For a fixed telescope, this is proportional 
to $\sqrt{AZ\xi T}=\sqrt{ST}$ where $A$ is the area of the telescope, $Z$ is 
the source strength, $\xi$ is the efficiency of the telescope and detector, 
and $T$ is the observing time. Other sources of 
noise can reduce the signal-to-noise ratio. The ratio of the Poisson variance 
to the real variance is a measure of the efficiency of the readout scheme. 
Variance is used rather than sigma, so the efficiency relates directly to time.
In what follows, we will use this efficiency to compare various strategies.

In IR detectors, the photons cannot yet be reliably counted individually. The 
detector and readout electronics add noise
(Fanson \etal\ 1998, Tian \etal\ 1996). This can be summarized as the 
readout noise, $R$, expressed as a number of photons or electrons. Typical 
values range from a few to $\sim40$~e for well-designed systems. A 
nondestructive readout allows the use of multiple samples on the same pixel and 
electrons to reduce the readout error. However, the additional samples make a 
more complex system with more electronics and computational complexity. The 
process of sampling the detector can add electronic noise directly, or heat 
the detector, causing additional noise. With the current state of the art
detectors, the noise can be reduced by increasing the number of samples.

There are often sources of noise with a power spectrum that scales inversely
with frequency. Although this noise is poorly understood, it is often
present, and in many detectors it becomes the dominant source of noise at
very low frequencies. This sort of noise
increases with time just like the photon noise from a source, so it can be 
treated like a background (or foreground) contamination.

The readout noise decreases with the number of samples, and the signal
(and Poisson variance) increases with time. Thus after a sufficient time and 
number of samples, the readout noise will be insignificant relative to the 
photon noise, and the ideal efficiency of $\epsilon=1$ will be approached.

However, there are two obstacles to long integration times. At some point,
the wells of the detector fill up, and the detectors become nonlinear and 
ultimately insensitive to additional photons. Also, particularly in space 
based detectors, cosmic rays strike the detector at a rate of 5-30 
cm$^{-2}$sec$^{-1}$ (Tribble 1995, Barth \& Isaacs 1999) (much higher in 
radiation belts or during solar flares) and add a large noise signal.

\section{Sampling Strategy} 
The sampling strategy is an integral part of the larger observation strategy.
The best strategy depends, of course, on the goal of the observation. Possible 
goals include finding sources, surveying a region of the sky, identifying
sources, examining a source for variation, or mapping diffuse emission.
Here we concentrate on detecting dim sources and assume that band
limiting filters are used for purposes of identification and analysis.

If only a fixed number of samples, $n$, are available, there is an advantage of 
using them at the beginning and the end of the integration (Fowler sampling,
Fowler \& Gatley 1990, and Fowler \etal\ 1996). This approach gives the longest 
effective integration time, and the variance from the readout noise is reduced 
by $4/n$, if the sample noise is uncorrelated. Half of the samples are used at 
the beginning and half at the end, and the signal is obtained by differencing, 
which adds the variance from the beginning and the end resulting in the 
factor of 4. The efficiency is then $\epsilon=ST/(ST+ 4R^2/n)=1/[1+4R^2/(nST)]$, 
which approaches 1 for long times $T$, large signals $S$ or many samples $n$.
This does not account for cosmic ray hits or other sources of noise which
also have effects.

A fixed sample rate has advantages, as the electronics can be made simple, and 
the effects of the readout on the chip (heating, electronic noise) can at 
least be made constant. Also, the detector charge history can be searched for 
cosmic ray strikes and other glitches. Uniformly weighting each of the readouts
and fitting to a straight line effectively uses only 2/3 of the 
integration time (difference between the mean negative weighted times and the 
mean positive weighted times).
Hence, the maximum efficiency is .667 for dim sources. For bright sources, as 
we shall see in section 5, this limitation can be overcome, and the 
efficiency can approach 1.

In the limit where the variance is dominated by the readout noise, it is 
$V=12R^2n/(n+1)/(n+2)$ so the efficiency, $\epsilon=ST/[ST+12R^2n/((n+1)(n+2))]$
is only 1/3 of the Fowler readout.
But increasing the integration time has a dramatic impact because it increases
the number of samples and the signal, so the signal-to-noise ratio increases 
as $T$ for read noise dominated uniform sampling.

\section{Hardware Considerations}
Hardware limitations often run counter to the algorithmic requirements.
For instance, in order to make Fowler sampling efficient, it is desirable
to have rapid sampling at the beginning and at the end of the integration.
But the high frequencies required for high speed sampling add more noise, and
A$\rightarrow$D converters generally have more noise at higher speed. Ultimately, this 
limitation is related to the noise voltage on a resistor, which is proportional 
to the square root of the bandwidth, which is inversely related to the sampling time. 

Multiple readouts on each detector chip can relax the constraint, but only at 
the cost of additional wires, readout electronics and complexity. Also, each
readout path can have its own offset and gain requiring additional parameters
in the data processing.

The sheer volume of data can present a problem. An 8K$~\times~$8K detector 
running at a sample rate of 1~Hz produces 6~TB in just 12.4 hours of observation.
Storing and processing this 
volume of data daily requires a dedicated computer system.

\section{Processing}
Fowler sampled data are easy to deal with. Ideally, they consist of $n/2$ 
samples at the beginning of an integration and $n/2$ samples at the end of the 
integration. The data from the beginning and end are summed independently, and
the final result is the difference between the two. For integrations that are 
of the same order as the sample time, the readout sequence ends up taking a 
substantial fraction of the total time. Since the mean time of the readout
is the effective time, the effective integration time is reduced to 
$T-nt/2$, where $t$ is the time for a 
single sample. This reduces the efficiency to $1-nt/2T$ even if the 
readout noise is completely negligible compared to the signal.
The readout noise must be added, of course, but if the number of 
samples is large relative to $R^2$, where $R$ is the readout noise in photons 
(or its equivalent electrons), the readout noise becomes small, and any signal 
is dominated by the photon noise. 

For a typical size detector pixel ($600~\mu$m$^2$), the mean time between 
cosmic ray events is $\sim40000$~sec in space. The problem with a cosmic ray 
is that {\it all} of the information is lost in the affected pixel due to the 
cosmic ray, and the efficiency drops to zero. So the integration time is a 
balance between making longer integrations to reduce the readout noise effect 
and losing information to cosmic rays. Also, if a pixel well is filled, 
the information is lost.

For a data set formed by uniformly sampling with non-destructive reads,
the important signal is the time derivative or slope.
Finding the slope of a set of data can be done in a 
straightforward manner. But there are several important sources of noise,
and if attention is paid to each of these, the process can be optimized as 
shown in $\S$7.
Full wells are recognized, and data taken after a well is filled are
ignored. This reduces the effect on the pixel from losing all of the 
information to effectively losing only the time after the well is full.
Cosmic ray glitches are identified and eliminated from the data
before fitting the slope. Finally, the weighting is adjusted to weight
high signal pixels towards the ends of the integration and low signal pixels
uniformly. This optimizes the fitting so there is no penalty for the
high signal data over the Fowler sampled data. Thus, if the integration
is carried out long enough, the readout noise can be effectively removed for
even the low signal data, although this will suffer a factor of $\sqrt 3$
noise increase over Fowler sampled data. But if the integration time
is a factor of 2 or more longer, uniform sampled data have a higher efficiency.

Many other strategies for sampling can be devised. However, they tend to
combine the disadvantage of the hardware complexity of Fowler sampling
with the disadvantage of the software complexity of uniform sampling.
Unless there are peculiar features of the hardware or noise, they offer
little improvement over the two cases presented here.

In Figure 1 we compare the efficiency of uniform sampling with optimum 
processing to Fowler sampling with its simpler optimum processing.
A range of brightnesses is presented along with a range of integration times.
We assume a cosmic ray rate of $25\times10^{-6}$~s$^{-1}$ per pixel 
(5~cm$^{-2}$~s$^{-1}$ for 27~\um\ detectors 10~\um\ thick) and a full well of 
70000 e. This introduces the sudden drop to zero in the Fowler sampling 
and the sharp decrease in the uniform sampling for the brightest sources. 
We also assume a minimum sample time of 10 seconds,
and a maximum number of samples of 64 per integration. This means that for
integrations less than 640 seconds, readout is continuous and uniform even for
the Fowler case. The uniform curves are higher because the data
are optimally processed, while the identical Fowler data are Fowler processed.
This limits the Fowler efficiency to 50\%. For longer times, it can be seen that
the Fowler sampled data have a peak at $\sim 2000$~sec for bright sources and
up to 5000~sec for dim sources. The fall-off is due to cosmic ray hits.
Changes in the size or rate of cosmic ray hits change the exact placement
of the peaks but not their character.

The uniformly sampled data also suffer from cosmic ray hits, but since they are
removed by the processing, their effect is small, and efficiency continues to 
increase until the wells are filled. Thus, the efficiency of the uniform 
sampling can exceed that of the Fowler case provided a sufficiently long 
integration time is used. As noted above, this is approximately a factor of 
2 for low level sources.

\section{Data Handling}
Moore's law (memory doubles every 18 months, Moore 1965) suggests that we 
should not have to worry about data storage (at least not for long), but 
detector arrays also double on the same time scale, and the problems of 
data storage and handling
remain. Keeping only the appropriate bits can substantially reduce the data 
volume. Obviously, any large offset can be removed, and the size of the data 
can be scaled to the largest measurements. Excessive precision includes noise
that is not useful and cannot be compressed. Insufficient precision allows
the uncertainties to be dominated by the digitization rather than the 
original measurement. Making the least significant bit the size of the 
uncertainty, ($1\sigma$), increases the variance by 8\%, an additional
bit reduces the increase to 2\%, and 2 extra bits results in only a .5\%
variance increase. Thus we arrive at the well-known injunction to keep one 
digit into the noise but no more.

If the noise of a system is constant, the A$\rightarrow$D can be arranged so that the
$1\sigma$ noise is $a$ times as large as the least significant bit. The value 
of $a$ is then typically chosen to be in the range 1-10. The precise value 
depends on the relative cost of keeping extra bits of marginal value to losing 
a small part of the signal to noise ratio.
If the noise of the system is variable, the process is more difficult. The
A$\rightarrow$D can be adjusted for the lowest noise data. The data can then be divided
by the noise, multiplied by $a$, and the integer result stored. If each datum
has a unique noise, a precise noise estimate must be stored for each one,
again increasing the number of bits that must be stored.

Poisson distributed data present an efficient alternative. Since the uncertainty
is $\sqrt N$, dividing by the uncertainty gives $\sqrt N$, and both the signal
and the noise have the same record, $a\sqrt N$. Hence, there is no need for a 
separate noise file, and the data can be restored to sufficient accuracy simply 
by dividing by $a$ and squaring the result. Although we have looked at this 
problem in the context of astronomical observations, this particular data
compression technique is general and can be used anywhere the variance
is proportional to the signal.
 
In the case under consideration, the noise is not purely Poisson noise, but the
variance can be well approximated by $N+V$, where $V$ is related to the readout
noise and the number of samples. Since we are dealing with a variation on a
small perturbation of the noise, a reasonable approximation will work.
After compression, neighboring values will be combined, by truncating any 
residual fraction and keeping only the integer part. The key is the range of
values that will be combined, not the absolute value which can be recalculated.
Consider the form 
\begin{equation}
D=a\sqrt{N+f},
\end{equation}
where $a$ is the scaling factor as before, and $f$ is an offset. Now for large
signals $D\approx a\sqrt N$ as desired. For small signals:
\begin{equation}
a/\sqrt V=dD/dN=a/(2\sqrt{N+f})
\end{equation}
which is satisfied at $N=0$ for $f=V/4$. The encoding is fast, requiring only
an add, square root and multiply. The decoding is also fast and straight-forward.
This process reduces the number of bits by a factor of 2 with essentially no
loss of information. The advantage remains even after additional lossless
compression (Nieto-Santisteban et al. 1999).

\section{Algorithm}
The algorithm to process Fowler sampled data has already been discussed. We now
turn to the problem of optimizing an algorithm to process uniformly sampled
data. All of the sources of error must be considered to make this a general
algorithm. This will unnecessarily complicate it for some cases, but the
computer implementation can take advantage of special cases. 

The algorithm will assume a nondestructive set of samples for an integration.
We will denote the raw data $P_i$, and assume $n+1$ samples at times [0,1,...n].
Thus, the index number is the time, and we look for a signal per sample time.
We will also frequently use data differences, $D_i=P_i-P_{i-1}$, which 
approximate the signal.

The first issue to consider is full wells. Conceptually, the detector collects
electrons until the well is full and then stops. Real detectors often show
some nonlinearity before the well is full, and may show some ``bleeding" into
neighboring pixels. In any case, each pixel's data must be examined to 
determine if the pixel's well is full. The examination is most efficient if
it proceeds from the last datum towards the first and stops when the first
sample less than full is reached. By using all of the data up to the full well,
maximum dynamic range is maintained. The limit of dynamic range is then 
determined by the sampling frequency and the full well size. To account for 
residual nonlinearity, it is most time efficient to use a look-up table to 
provide a linearization function.

Next, the data must be searched for cosmic rays and glitches (glitches of 
either sign might arise from cosmic ray strikes on electronic elements other 
than the detector, or other sources). Since glitches can appear at any time, all
of the data must be examined and compared to what is expected. So, we must
make an estimate of what is to be expected. A simple and robust signal estimate 
is the difference between the last and first data divided by the time: 
$s=(P_n-P_0)/n$. A difference between $s$ and $D_i$ should be within the
noise. One might object that $D_i$ is included in $s$ as $s=\sum D_i/n$,
but a little algebra shows that this only reduces the difference by a factor
of $(n-1)/n$. This can be ignored for large $n$ or included in the expected
uncertainty. The expected uncertainty must include 2 readout uncertainties,
as $D_i$ is a difference, and the Poisson statistics of the photons. But
a large cosmic ray hit included in the estimate of $s$ could perturb it to
the point that other $D_i$ are interpreted as outliers. The solution to this
problem is to identify the {\it worst} offender. If it is within bounds, then
all of the other data are as well. If it is not, it must be discarded and 
excluded from the estimate of $s$. The test is then repeated on the remaining
data. There remains the question of how to set the bound. We leave this as a 
tunable parameter to adapt the algorithm to different cosmic ray rates, 
detector characteristics and observation goals.

Thus, $s$ is a good estimate of the signal. However, if the readout noise is 
the dominant source of noise, a linear fit to the data can reduce the variance 
by $12n/[(n+1)(n+2)]$. If, on the other hand, the noise is dominated by the 
Poisson statistics of the photons, the linear fit effectively uses only 2/3
of the time. In that case, the noise is worse for the linear fit by 23\%.

An optimum weighting is needed. The set of differences, $D_i$, has all of the
information. The covariance matrix for the set of differences can be expressed
as the sum of two matrices. Since the photon noise for each difference is
independent of each of the other differences, the matrix describing the
covariance due to the photon noise is $sI$, where $s$ is the signal and $I$
is the identity matrix. Neighboring sample differences are correlated by the 
readout noise. Each sample, $R_i$, with its noise, is used twice, once positive
(in $D_i$) and once negative (in $D_{i+1}$). Thus, the readout covariance
matrix has a main diagonal of $2R^2$ with a diagonal of $-R^2$ on either
side of the main diagonal. The sum of these two matrices can be inverted
to form a weight matrix. Since each difference is an estimate of the 
signal, the sums of the columns (or rows as the weight matrix is symmetric)
form a weight vector that optimizes the signal when applied to the differences.
By subtracting the weight vector from the weight vector shifted by 1, the
difference operation is effectively transferred to the weight vector, and the 
result can be applied directly to the raw data.

The final weight vector depends on the number of data elements and the 
signal-to-noise ratio $s/R^2$. For large $s/R^2$, only the two end points are 
used, and for small $s/R^2$ the weights have a linear slope with mean of 0
(see Figure 2).
Relatively few cases of $s/R^2$ can approximate the ideal full matrix
calculation quite well. In the algorithm shown below, we use 8 approximations,
and select the appropriate one by using the previously calculated $s$.
Selecting a poor weight vector does not bias the result, it only increases
the noise (and noise estimate) of the result. The error is not symmetric.
Using a weight for high signal-to-noise ratio for a low signal-to-noise ratio
case can result in an increase of order $n$ in the variance while an error in the other
direction at most increases the variance by 33\%. So we calculate the 
weight vectors for the bottom of each bin, see Figure 3.

\section{Non-Ideal Detectors}

All detectors are expected to be non-ideal, particularly after exposure to a 
space radiation environment (The NICMOS Data handbook). The algorithm described 
above was developed to demonstrate the advantages of uniform 
sampling, but it does not directly address detector faults. The 
strategy for accounting for each fault depends on its type.

Each pixel may have a different noise level and responsivity, and 
these parameters may change with time, particularly after a cosmic 
ray event depositing large amounts of charge. If a single cosmic ray 
can produce a responsivity change, then the algorithm would have to 
test for that effect and decide whether data taken after the cosmic 
ray are consistent with those (presumably correct) data before it. 
If the noise level can change with time, then the threshold 
criterion for recognizing a glitch may need to be adaptive rather 
than using a stored constant. However, neither of these cases is 
expected to raise the computation time very much, since most of the 
time is still devoted to input.

Many detectors show a distinct initial transient after 
the reset of the integrators. If this effect has the same general 
shape for all pixels, it can be modeled and subtracted from the 
measured data before processing to search for glitches. Such a model 
might be implemented as a simple formula, or as a table lookup. If 
the effect differs significantly between pixels, then a more complex 
model or table lookup may be required. However, it seems unlikely 
that a completely separate model for every pixel would be required. 
Improper treatment of such transients could affect the photometric 
accuracy for pixels that have glitches or become saturated early. On 
the other hand, such data are lost altogether for Fowler sampling. 
Improper treatment could also lead to false alarms, e.g. glitches that are 
the result only of the repeatable initial transient.

It may also be that each cosmic ray produces a temporary response 
that disqualifies data taken shortly afterwards. If this effect can 
be described by a simple dead time, it is easily handled in the code. 
If there is a long term effect with a constant waveform, it can be 
corrected by modeling and subtracting it. If it is still too difficult 
to correct, then data taken after the glitch can be ignored until the 
next exposure. The uniform sampling with cosmic ray rejection would 
then have a lower efficiency, but still would be superior to Fowler 
sampling in which such pixels can not be used at all.

The algorithm described here makes no use of spatial information. 
However, a processor could easily examine the neighbors 
of every pixel where a glitch or saturation has been found, to see if 
there is any odd behavior there. There are two problems: First, in this process
it is very easy to introduce spatial correlations which can contaminate the 
results. Second, the processing can become very involved making it slow. 
Either of these is incompatible with our attempt to produce
a simple, fast and robust process to remove CR and compress data from
large-area detectors.

Some faults might not be in the detectors. A variable pointing error 
would produce a variable brightness incident on pixels where spatial 
gradients are high, leading to false detections of glitches. These 
could be handled by raising the threshold for glitch detection for 
bright pixels, possibly by using an adaptive calculation.

To summarize, detector faults will require additional code in the 
processing algorithm, but this additional code is not expected to produce 
large computational loads, nor to invalidate or bias the data. For long 
exposures, uniformly sampled data will still be valid over a larger 
fraction of the pixels than Fowler sampled data.

\section{Algorithm Code, Processing Time and Performance}
The process outlined here does not imply anything about neighboring pixels.
This reduces the likelihood of generating any biases in the data. It also
keeps the process simple and short. Only 6 operations per datum plus 6 
operations per pixel are required to check, fit and compress the data.
Rejected data are somewhat more expensive, but presumably they are relatively
rare, so the entire process can take place on a modest computer for even
enormous data flows.

The data from an array are usually collected in the order of columns, rows
and then sample iterations. The order of the data required for the algorithm
is by sample iteration first and then pixel (either columns or rows first). 
This has two implications. First, an entire read set must be stored in memory.
For example, in the Next Generation Space Telescope (NGST) with 64 reads in
an integration, this set is 8 GB (2 bytes $\times$ 64 reads $\times$ 
1 M pixel per chip $\times$ 64 chips, see The NGST Study Team, 1997), so the 
task is not trivial. 

Second, to speed the
memory operations, it is convenient to have the addresses for the computer
reads in a different order than for the A$\rightarrow$D converter writes. If an even
power of 2 is adopted for the number of samples, this is easily achieved by
swapping some of the address bits and using a double buffer so that the computer
can be processing a set of data while the next set is being collected.

Since the compression ratio is large (of order 200 is easily achievable for 64
samples), the output storage and transmission are greatly simplified. The input 
and output bounds on the computer are entirely dominated by the input side.

The algorithms introduced in Offenberg et al.\ (1999) and Nieto-Santisteban 
et al.\ (1999) have now been optimized to reduce computer requirements and improve 
performance. First, saturated data are marked and not used.
Next, for each pixel, the set of samples (64) is fit to a straight line. 
The interval with largest deviation from the line (either direction)
is compared with the expected noise. If it is larger than $4.5\sigma$
(optimum for test case), 
the interval is not used in the fit, and the process is repeated. Most 
of the processing time is used by the cosmic ray rejection. 

Next, a weighted fit is applied to the remaining data. The optimum fit 
depends on the signal. High signal uncertainties are dominated by photon 
(electron) counting noise, and the optimum fit weights the endpoints. 
Low signal uncertainties are dominated by readout noise, and the optimum fit 
is uniform weighting. We calculate the weights for these and 6 intermediate 
signal-to-noise ratios and choose the best weighting scheme for the signal. By 
computing the weight table for all 8 signal-to-noise ratio levels ($<1\%$ 
of total computation time) and all 
possible segment lengths, we save time in the weighted fit.

After the fit, we reduce the dynamic range and equalize the noise for the 
different pixels by finding the square root of the slope plus an offset, 
which compensates for the readout noise. Finally, an adjustable scaling 
allows retention of $a$ bits of noise after conversion to an integer. Thus N 
(64) 16 bit samples are converted to a single 8 bit byte. This can be further 
compressed without loss (see Nieto-Santisteban et.al. 1999) to approximately 3 
bits per pixel (if we keep 2 bits of noise).

The final results are robust. Even integration times that lead to most of
the pixels being affected by cosmic rays can be effectively cleaned,
allowing longer integration times than are practical with Fowler sampling. 
This leads to higher efficiency than Fowler sampling. The difference can
be as large as a factor of 2, which translates directly to saving half of the
total integration time while getting the same quality data.

Figure 4 demonstrates the effectiveness of cosmic ray rejection. The cosmic 
rays in the raw image (or, equivalently, in a Fowler-sampled image) make long
integration ($>1000$ sec) undesirable. If we take 
non-destructive samples every 30 seconds during the integration, it is 
possible to get a clean image of long integration times, with high photometric 
reliability. The images shown in Figure 1 are stretched to the same 
grey-scale and assume a cosmic ray rate of 5 event/sec/cm$^2$, which is 
the low end estimate for cosmic ray rates in deep space (L2 orbit), such as
planned for the Next Generation Space Telescope. 

The lower pair of plots in Figure 5 compares photometry for Fowler sampling and 
uniform sampling with processing for five 2000 second integrations. 
The resulting images are processed by throwing out the outliers and using IRAF 
DAOPHOT on the final image. The Fowler sampled data are still contaminated
with cosmic rays (CR), as expected, so the ideal integration time is shorter. 
But the ideal integration time is longer for uniform
sampling with deglitching. The upper pair of plots in Figure 5 compare 
the optimum twenty 500 sec Fowler sampled images with a single 10000 sec uniform 
sampled image after deglitching. In all cases, a total of 320 samples
were executed for 10000 seconds of observation. The uniform sampled 
image is higher quality even though the data rate is only 1/40 as large 
(after data compression). 

As Table 1 indicates, even though the {\it processing} (including CR rejection)
of this algorithm takes longer than the {\it processing} of Fowler sampled 
data, the total time is dominated by I/O and in particular input. If the input 
speed can be raised to the data rate, the remaining
processing for either the Fowler sampling or this algorithm is
accommodated on a modest computer.

\begin{table}
\caption{Uniform w/CR vs. Fowler}
\begin{tabular}{rcc} \hline \hline 
&CR Processed &Fowler Processed \\
 \hline
Total input:~~~    &10.7 GB    &10.7 GB \\ 
Max data rate:~~~  &10 Mb/sec   &100 Mb/sec  \\
Input time:~~~     &2600 sec (81 op/Dat) & 2600 sec (81 op/Dat) \\ 
Process time:~~~   &1450 sec (45 op/Dat)  &200 sec (6 op/Dat)  \\ 
Convert to float:  &400 sec (12 op/Dat) & - \\
CR Identification: &750 sec (23 op/Dat) & - \\ 
Weighted Fit:      &400 sec (12 op/Dat) & - \\
Compression time:  &50 sec (2 op/Dat)  & - \\
Output time:~~~    &100 sec (3 op/Dat)     & 250 sec (8 op/Dat) \\
Total output:~~~   &50 MB          &128 MB \\
Total time:~~~     &4200 sec (131 op/Dat)       &3050 sec (95 op/Dat) \\
		\hline \end{tabular}
			\protect\label{table:summary}

Note: The times are based on a Ultra Sparc 2 running at 168 MHz.
\end{table}

\section{Conclusions}
The algorithm presented here is robust as it excises points far from the line
and uses a linear fit on the points near the line.
It preserves the maximum dynamic range allowed by the hardware readout
and rejects almost all cosmic ray hits. Its adjustable weighting efficiently
uses uniformly sampled data, and yet it uses a minimum of computer resources.

These studies are supported by the NASA Remote Exploration and
Experimentation Project (REE), which is administered at the Jet
Propulsion Laboratory under Dr. Robert Ferraro, Project Manager.

\appendix{}
\section{Cosmic Ray Rejection Program}
The following is the complete algorithm. The program "init\_wt", is required 
only once to set up the weight table, WT.
The output of the program, "cr\_rej" is a compressed best fit image and
a list of the number of glitches found at each location.

\begin{verbatim}
#include <math.h>
#include "simpleimage.h"
#define e 2.71828
#define E .0024788                  //smallest S/N ratio bin
#define K 8                         //number of signal (S) cuts
#define M 64                        //Max number of reads
float WT[K][M][M/2+1];

void init_wt(float *a, float *f, float *v, float *q, 
float B,                //Bits of noise to keep
float P,                //Gain (photons/count)
float V,                //Readout variance (counts^2)
int N,                  //Number of reads
float S)                //Sigma for CR cutoff
{ float *W; double Z[M],Y[M],s,w,x,y,z; register int i,k,n;
  *a=B*sqrt(P);                                        	//Renormalize Bits 
  *f=3.*V/(N*P*(N+1)*(N+2));                            //Output offset
  *q=S/P;                                               //Renormalize cutoff
  *v=2.0* *q*V*(N*N+1)/(P*N*(N-1));                     //Renormalize variance
  for(s=E,k=0;k<K;s*=e,k++){WT[k][0][0]=0;              //loop on S/N ratio, s.
    for(Z[0]=y=x=n=0;++n<N;){W=WT[k][n];                //for all ramp lengths
      Y[n]=x=1/(2+s-x);y++;                             //New x New last Y
      for(z=i=0;++i<n;)z+=Z[i]+=y*(Y[i]*=x);            //Update row and sum
      for(y=x,i=0;++i<n;)y+=Y[i];                       //Sum weight
      for(i=0;i<(n+1)/2;i++)W[i]=(Z[i]-Z[i+1])*P*(N-1); //Differences
      W[i]= z+(Z[n]=y);}                                //final weight
    for(x=1/(z+y),i=0;i<N/2;i++)W[i]*=x;}}              //Renormalize last row

void cr_rej(//Routine to reject cosmic rays and perform linear fit of data.
short int *DI,          //Input: Data cube (np x N) containing the data.
const int N,            //Input: number of images in the data set.
simpleimage *CD,        //Output: contains the fitted image data.
simpleimage *CR,        //Output: number of detected CR
const int np,           //Input: number of pixels in the data set.
const int F,            //Input: count for Full-well.
const float a,          //Input: noise bits to keep in sqrt-scaling
const float f,          //Input: offset for sqrt-scaling
const float v,          //Input: renormalized read variance 
const float V,          //Input: Read variance (counts^2)
const float q)          //Input: Chi^2 limit in counts
{ register int p,c,k; register short int *D, *P;
  const float EVe=E*V*e; const int n=N-1;
  register float s,x,y,*U,*W,*X,*Y,Z[M],*C[M]; U=Z+n;
  for(P=DI+N*(p=np);p-->0;){D=P;P-=N;                   //Do for all pixels
    if(P[1]>F){CR->setval(p,N);CD->setval(p,0);}        //If no good data exit
    else{while(*--D>F);                                	//Find last good data
      for(c=-1,X=Y=Z+(k=D-P);(C[++c]=Y++)<U;);          //Delete Full Reads
      for(Y=X;D>=P;*Y--=*D--);                          //Convert to floats
      s=(*X-*Z)/k;                                      //Estimate Signal
      for(x=v,Y=X;Y>Z;)if((y*=y=s-*Y+*--Y)>x){X=Y;x=y;} //Find Worst Point
      for(;x>s*q+v;X--){                                //Is it a CR?
        s+=(s+*X-*(X+1))/(n-++c);                       //fix s for CR
        for(k=c;((C[k]=C[k-1])>X)&&--k;);C[k]=X;        //fit CR in list
        for(k=0,x=v,Y=Z;k<=c;++Y){W=C[k++];             //for all segments
          while(Y<W)if((y*=y=s+*Y-*++Y)>x){X=Y;x=y;}}}  //Find Worst Point 
      CR->setval(p,c);                                  //Record # CR
      for(k=0,x=EVe;s>x&&++k<K-1;x*=e);                 //Find proper S/N Bin
      if(c){for(Y=Z,s=y=c=0;Y<U;Y=C[c++]+1){            //for all segments
          W=WT[k][C[c]-Y];                              //Find right weights 
          for(X=C[c];Y<X;)s+=*W++*(*Y++-*X--);          //total this segment
          y+=*W;}                                       //Sum of weight
        s=s/y+f;}                                       //Final Sig
      else for(W=WT[k][n],s=f,X=U;*W<0;s+=*W++*(*Y++-*X--));//0CR weighted fit
      CD->setval(p,(s<0)?0:int(a*sqrt(s)));}}}          //Record data
\end{verbatim}

\begin{figure}
\label{eff}
\caption{The efficiency of uniform sampling with optimum processing
(solid lines) is compared with Fowler sampling (dashed lines) processed with
a simple difference as a function of integration time. A cosmic ray rate of 
5~cm$^{-2}$~s${^-1}$, and a full well of 70000 e are assumed. The efficiency is 
higher for brighter objects so plots are shown for .01 .1, 1, 10 and 100 e/s. 
We assume the sample rate is limited to .1 Hz for the array (100 kHz sample 
rate) so for short times the sampling is identical for the Fowler and uniform 
cases. The optimum processing gives the uniform case an efficiency advantage 
here. For longer integrations we assume a maximum of 64 samples over the 
integration time.}
\epsffile{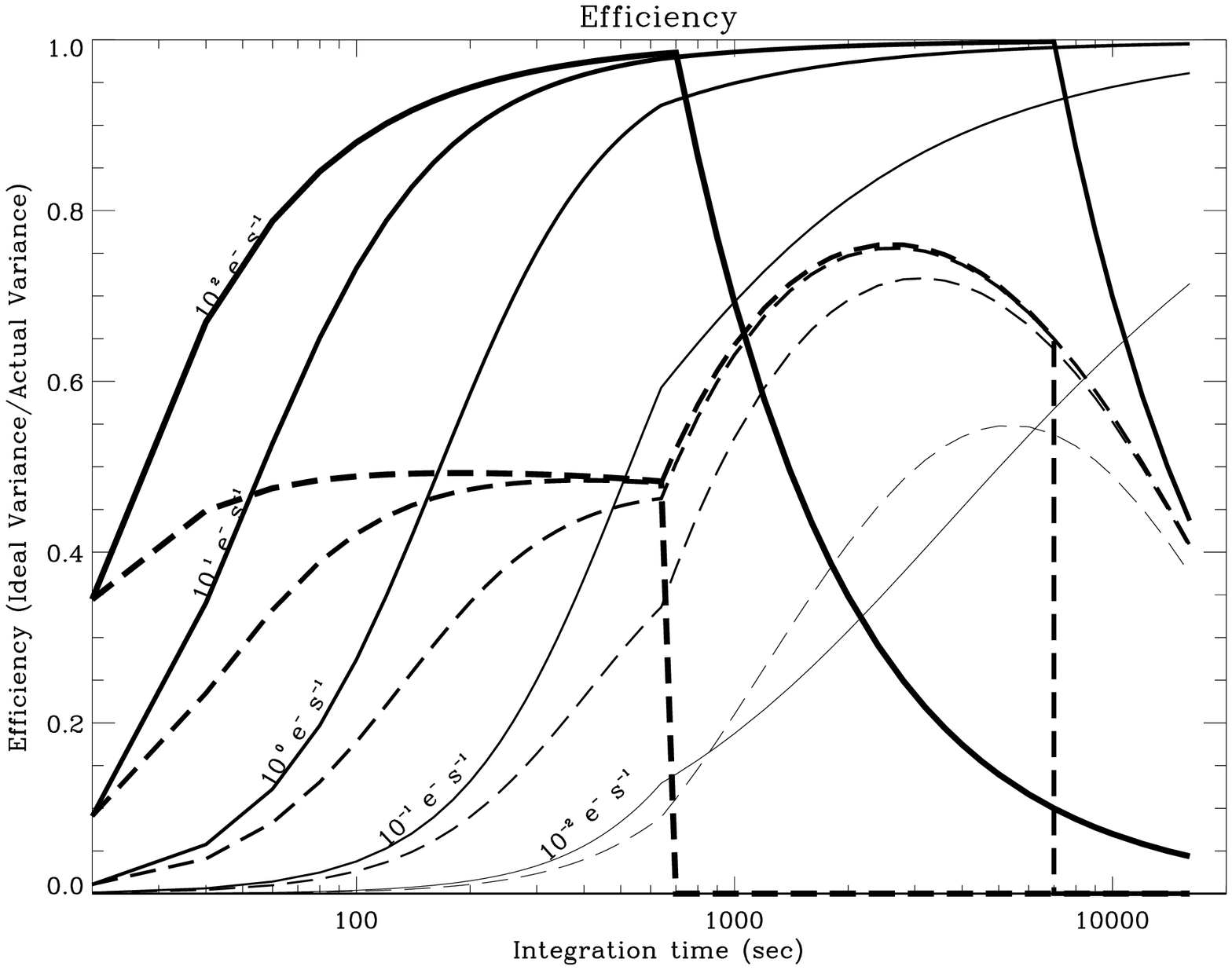}
\end{figure}

\begin{figure}
\label{2D}
\caption{The weights for each of the eight bins are plotted as a function of 
time. The weights for brighter (higher signal-to-noise ratios) pixels are
thicker lines. The limit for zero signal is a straight line. The limit for
large signals is zero weight everywhere except the two end points.}
\epsffile{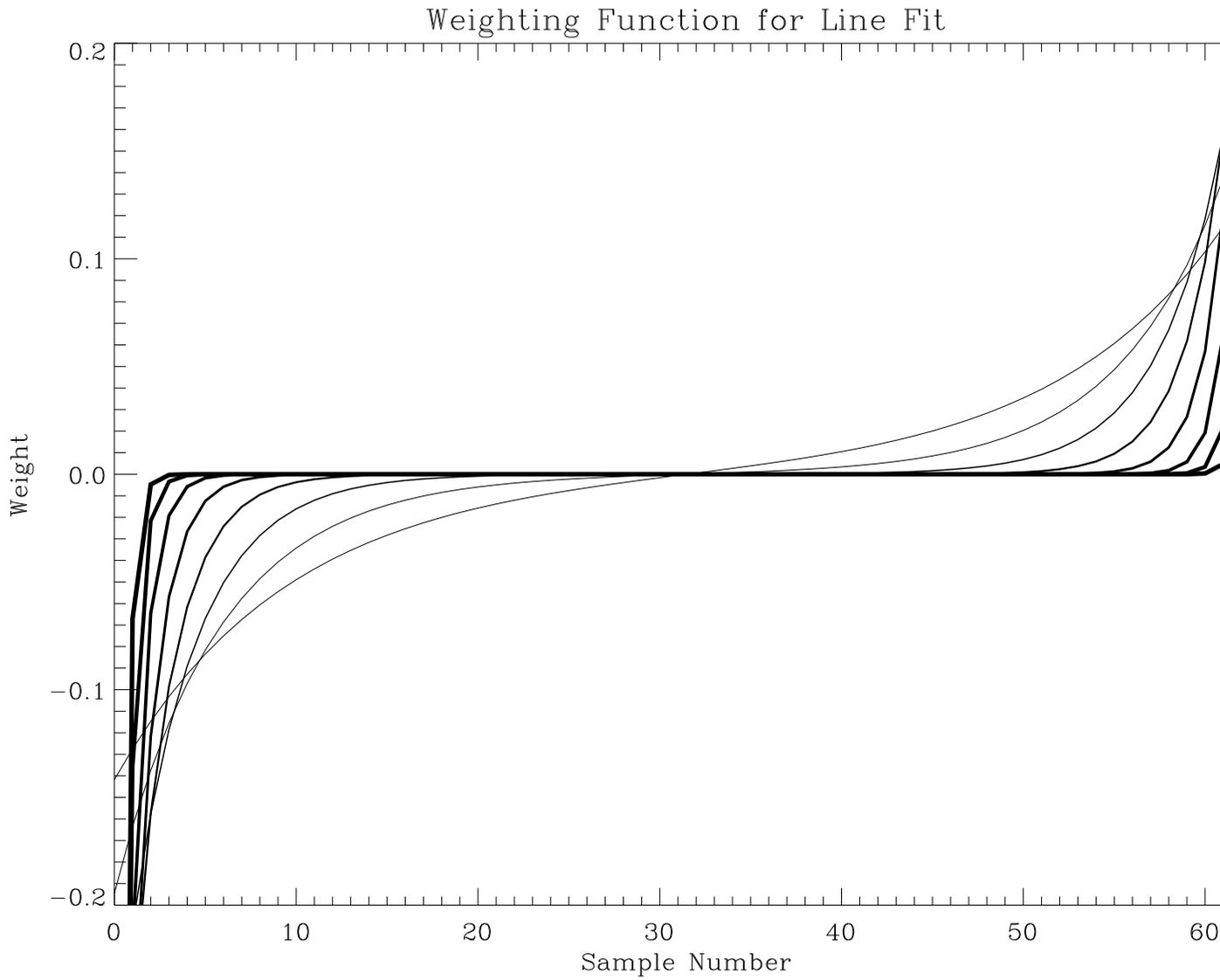}
\end{figure}

\begin{figure}
\label{3D}
\caption{The optimum variance/variance is calculated as a function of the 
the true signal-to-noise ratio and the assumed ratio. The result is far from
symmetric. It is safer to assume a low signal-to-noise ratio. The steps of the
approximation used in the algorithm are plotted as the solid line.}
\epsffile{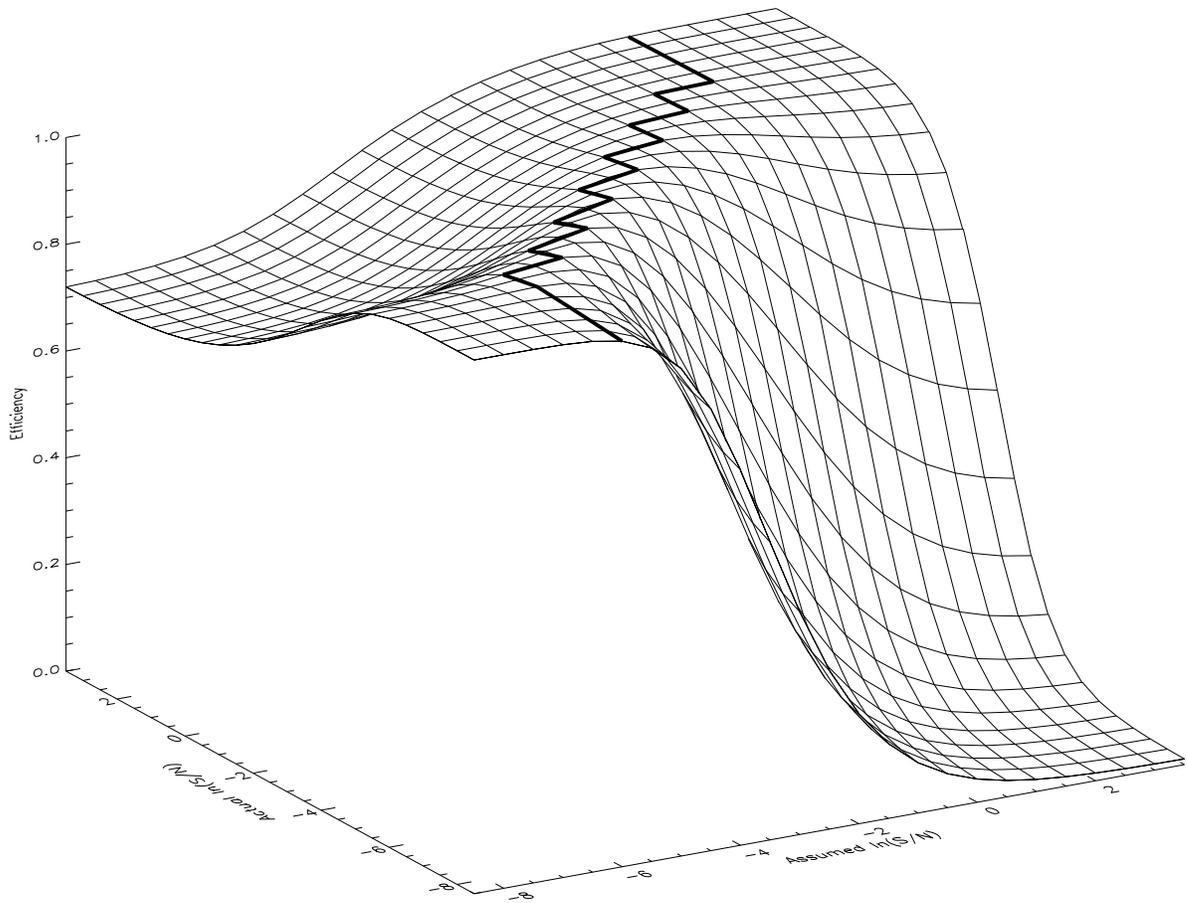}
\end{figure}

\begin{figure}
\label{Picture}
\caption{The image at left is a simulated image after a 10000 sec 
integration. It corresponds to an ideal Fowler processed result. The image
on the right is the result of processing and compressing the data after
uncompression. With 63 intermediate samples the cosmic rays can be effectively
removed.}
\epsffile{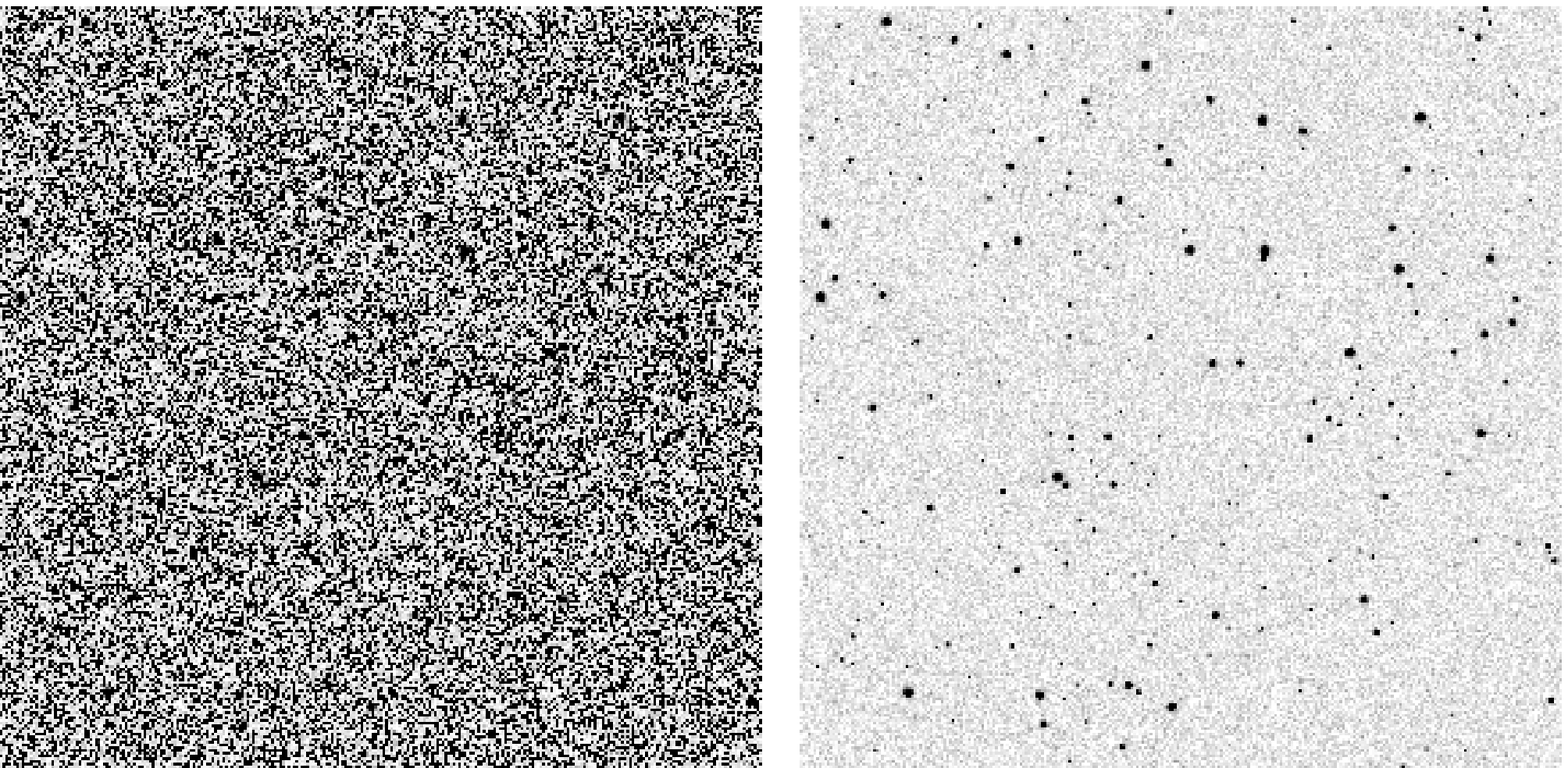}
\end{figure}

\begin{figure}
\label{}
\caption{Plots of photometric errors are shown for Fowler and uniform
sampling. In all cases a total of 10000 sec of integration were used. The
uniform case is shown for a single integration and five 2000 sec integrations.
The Fowler case is shown for five 2000 sec integrations and twenty 500 sec
integrations.The data for multiple integrations are averaged after rejecting
outliers $>2.5\sigma$.}
\epsffile{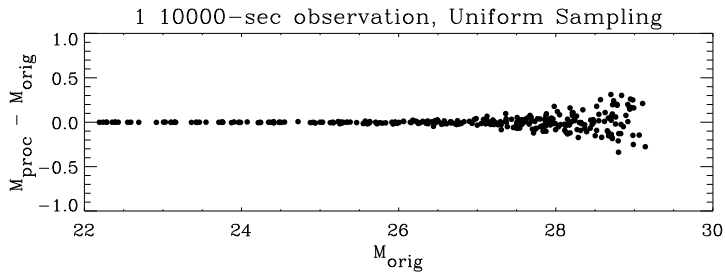}
\end{figure}

\end{document}